\documentclass[runningheads]{llncs}
\usepackage{amssymb}

\usepackage{float}
\usepackage[table,xcdraw]{xcolor}

\usepackage{lscape}
\usepackage[figuresright]{rotating}
\usepackage{tabularx}
\usepackage{booktabs}

\usepackage{graphicx}
\usepackage{physics}
\usepackage{amsmath}
\usepackage{tikz}
\usepackage{mathdots}
\usepackage{yhmath}
\usepackage{cancel}
\usepackage{color}
\usepackage{siunitx}
\usepackage{array}
\usepackage{multirow}
\usepackage{amssymb}
\usepackage{gensymb}
\usepackage{tabularx}
\usepackage{extarrows}
\usepackage{booktabs}
\usetikzlibrary{fadings}
\usetikzlibrary{patterns}
\usetikzlibrary{shadows.blur}
\usetikzlibrary{shapes}
\usepackage{diagbox}
\usepackage{bbding}

\begin{document}

\newtheorem{defn}{Definition}

\newcolumntype{C}{>{\centering\arraybackslash}X}
\title{Cross-user activity recognition via temporal relation optimal transport}
%
%
\author{Xiaozhou Ye\inst{1}\orcidID{0000-0002-9725-1548} \and
Kevin I-Kai Wang\inst{2}\orcidID{0000-0001-8450-2558}}
\authorrunning{X. Ye et al.}
%
\institute{Department of Electrical, Computer, and Software Engineering, The University of Auckland, Auckland, New Zealand \email{xye685@aucklanduni.ac.nz}\\
\and Department of Electrical, Computer, and Software Engineering, The University of Auckland, Auckland, New Zealand \email{kevin.wang@auckland.ac.nz }}
\maketitle              
\begin{abstract}
Current research on human activity recognition (HAR) mainly assumes that training and testing data are drawn from the same distribution to achieve a generalised model, which means all the data are considered to be independent and identically distributed $\displaystyle (i.i.d.) $. In many real-world applications, this assumption does not hold, and collected training and target testing datasets have non-uniform distribution, such as in the case of cross-user HAR. Domain adaptation is a promising approach for cross-user HAR tasks. Existing domain adaptation works based on the assumption that samples in each domain are $\displaystyle i.i.d. $ and do not consider the knowledge of temporal relation hidden in time series data for aligning data distribution. This strong assumption of $\displaystyle i.i.d. $ may not be suitable for time series-related domain adaptation methods because the samples formed by time series segmentation and feature extraction techniques are only coarse approximations to $\displaystyle i.i.d. $ assumption in each domain. In this paper, we propose the temporal relation
optimal transport (TROT) method to utilise temporal relation and relax the $\displaystyle i.i.d. $ assumption for the samples in each domain for accurate and efficient knowledge transfer. We obtain the temporal relation representation and implement temporal relation alignment of activities via the Hidden Markov model (HMM) and optimal transport (OT) techniques. Besides, a new regularisation term that preserves temporal relation order information for an improved optimal transport mapping is proposed to enhance the domain adaptation performance. Comprehensive experiments are conducted on three public activity recognition datasets (i.e. OPPT, PAMAP2 and DSADS), demonstrating that TROT outperforms other state-of-the-art methods.

\end{abstract}

\section{Introduction}

Human activity recognition (HAR) is a crucial research area in Human-Computer Interaction, ubiquitous computing \cite{andreas2014tutorial} and the Internet of Things \cite{abdallah2018activity}, which aims to correctly classify the current activity of a human or a group of humans performs based on physical sensor observations and context information \cite{lara2012survey}. HAR is applied to a variety of applications, including medical treatment, ambient assisted living, fitness and sports, rehabilitation, security surveillance, health monitoring, automatic security and home automation.
Existing state-of-the-art HAR methods for processing time series sensor data are under the assumption that training and testing data are drawn from the same distribution, which means all the data are independent and identically distributed $\displaystyle i.i.d. $ \cite{wilson2020survey}. This assumption implies that a generalised model is achieved across the source training data and the target testing data. In other words, the training and testing data are taken from the same domain where the feature space and data distribution characteristics are the same \cite{pan2009survey}. However, in many real-world applications, this assumption does not hold, and the collected training and target testing datasets have non-uniform distribution due to the issue of data heterogeneity or sometimes referred to as the out-of-distribution ($\displaystyle o.o.d. $) problem. The performance of a model that is trained on the source domain will likely decline when tested on the target domain \cite{patel2015visual}. Therefore, it is necessary to consider a more practical scenario where the data distributions of the training and testing set can be different. 

Transfer learning (or more specifically domain adaptation) \cite{pan2009survey} is a promising method for handling data heterogeneity. In general, the principle and key idea of transfer learning is to find some common knowledge and reduce the data distribution differences between the source domain and target domain. Most of the current domain adaptation research focus on static data area such as computer vision with the assumption that each sample(image) is $\displaystyle i.i.d. $ in its corresponding domain \cite{patel2015visual}. The same assumption is also applied to time series transfer learning research, such as sensor-based HAR, which means each sliding window is $\displaystyle i.i.d. $ in its corresponding domain \cite{zhao2011cross}\cite{deng2014cross}\cite{chen2019cross}\cite{soleimani2021cross}. However, time series data typically have inherent temporal relation, which means each sliding window is not independent of each other \cite{esling2012time}. For an intuitive example of activity recognition, walking is made up of three sub-activities: raising the leg, thrusting forward and feet to the ground. These three sub-activities have temporal dependency relations. So, existing domain adaptation approaches do not fully utilize the characteristics of time series data effectively because temporal relation knowledge is overlooked in the process of aligning data distribution.

In this paper, we propose a novel temporal relation optimal transport (TROT) method targeting time series domain adaptation for cross-user activity recognition, which considers temporal relations across time. This method captures the data distributions of the common sub-activities while preserving the same temporal relation between the sub-activities across source and target users. Temporal relation preservation acts like a regularization for reducing the data distribution differences of the sub-activities across users. In this case, the data distribution mapping process can achieve better domain adaptation performance. Specifically, it is a combination of optimal transport and the Hidden Markov Model (HMM) method that performs cross-user HAR considering temporal relation knowledge. Moreover, we propose a new regularizer for the optimal transport domain adaptation problem that preserves temporal relation order information to ensure correct data distribution alignment. Comprehensive experiments on three public activity recognition datasets demonstrate the better performance of TROT in time series domain adaptation applications (e.g. cross-user HAR). The rest of the paper is organized as follows: Section 2 presents the related work, delving into cross-user HAR and the concepts of transfer learning and domain adaptation. Section 3 describes the proposed method, including problem formulation, the motivation behind it, and the specifics of the TROT method—focusing on temporal relation representation and OT-based data distribution alignment. In Section 4, we detail the experimental setup and provide a comparison between our method and existing methods. Finally, Section 5 offers conclusions drawn from the study and suggests potential avenues for future research.

\section{Related work}

\subsection{Cross-user human activity recognition}

HAR is one crucial part of ubiquitous computing because of its important role in supporting daily human life. It aims at recognizing and analysing human behaviours from learned high-level knowledge from multi-modal sensor observations and context information. Various types of sensors are utilized to implement activity recognition in different scenarios. From the sensor modality view, HAR can be classified into five types:  Smartphones/wearable sensors-based HAR, ambient sensors-based HAR, device-free sensors-based HAR, vision-based sensors HAR and other modality sensors-based HAR\cite{chen2021deep}\cite{lentzas2020non}\cite{wang2018device}. In this paper, we focus on wearable sensors-based HAR.

Sensor-based HAR is considered a problem of time series classification in machine learning research \cite{andreas2014tutorial}. A great number of classification models, such as ensemble learning \cite{sekiguchi2020ensemble}, SVM \cite{bulling2012multimodal}, and HMM \cite{amft2005detection} are proposed to solve the HAR problem. With the recent research development, deep learning has achieved many state-of-the-art results for a wide variety of tasks. Deep learning-based HAR approaches \cite{ding2018empirical} are able to learn high-level features and automatically extract features from massive amounts of data \cite{tan2018survey}. However, these approaches mostly depend on the assumption that training and testing data are drawn from the same distribution to ensure the generalization ability of the model, which means all the data are $\displaystyle i.i.d. $ \cite{wilson2020survey}. In many real-world applications, this assumption does not hold, and the collected training and testing datasets are $\displaystyle o.o.d. $. In this paper, we focus on the sensor-based HAR $\displaystyle o.o.d. $ problem.

There are various categories of sensor-based HAR $\displaystyle o.o.d. $: First, data is generated from different sensors. Different sensor types, platforms, manufacturers and modalities may cause different data formats and distributions \cite{xing2018enabling}. Second, the data pattern may change over time, which is also called concept drift \cite{lu2018learning}. For example, the walking pattern of the same person may be impacted by health status. Third, the behaviour differentiation between different people may also be significant \cite{saeedi2018personalized}. For instance, different people may walk at different paces. Forth, physical sensors may be installed in different body positions \cite{rokni2018autonomous} or environmental layouts in smart homes \cite{sukhija2019supervised}, which may lead to different data distribution. Our current research focuses on the sensor-based HAR $\displaystyle o.o.d. $ of behaviour differentiation between different people.

\subsection{Transfer learning and domain adaptation}

Transfer learning is a process that enables a model to be trained in one or more source domains and applied to one or more related target domains that have no label or few labels. During the process, one of the key tasks is to solve the $\displaystyle o.o.d. $ problem by reducing the distribution difference between source and target domains. Domain adaptation \cite{wilson2020survey} \cite{patel2015visual} is a special type of transfer learning. It still focuses on solving $\displaystyle o.o.d. $ problem between source and target domains but adds an additional constraint of assuming the same task of source and target domains. This paper focuses on the unsupervised domain adaptation problem, which means the availability of labelled data from the source domain and unlabelled data from the target domain \cite{wilson2020survey}.

Domain adaptation has been developed for many years, and most of the approaches are feature-based transfer learning \cite{pan2009survey}. The Large Margin Nearest Neighbour (LMNN) approach, as presented in \cite{weinberger2009distance}, utilizes Mahalanobis distance metric learning. The goal of this method is to learn a linear transformation, such that data from different classes are distinctly separated from each other within the feature space. Subspace Alignment(SA) \cite{fernando2013unsupervised} applied PCA to generate subspace for the source domain and target domain. Then, SA is achieved by minimizing Bregman matrix divergence to learn a linear transformation matrix. CORrelation Alignment(CORAL) \cite{sun2016return} learns a linear transformation by minimizing the distance between the second-order statistics of the source domain and target domain features. Optimal transport(OT) theory is a promising approach and is also applied to solve the domain adaptation problem \cite{cuturi2013sinkhorn} via Sinkhorn distances. There are several works extending the research of optimal transport domain adaptation. Optimal Transport for Domain Adaptation(OTDA) \cite{flamary2016optimal} is proposed to learn the coupling between two probability density functions. The source domain is then transformed into the target domain via barycentric mapping. Substructural optimal transport(SOT) \cite{lu2021cross} explores the substructure of domains to complete substructure-level mapping to achieve a balance between coarse-grained mapping and fine-grained mapping.

These feature-based domain adaptation methods focus on static data such as images, whereas time series data is often treated directly by the same domain adaptation framework \cite{lu2021cross}\cite{li2020simultaneous}. However, these existing domain adaptation approaches may not work well with time series data because the temporal relation knowledge from the time series data is overlooked in the process of aligning data distribution. Temporal relation is an essential characteristic of time series data. Existing approaches that only make use of temporal invariant knowledge lead to models with unreliable performance. This is especially the case in real-world cross-user HAR applications, where a model is trained by one (or more) user(s) and later applied to other user(s). Furthermore, temporal relation knowledge may be useful for finding the common knowledge between users for better sensor-based HAR domain adaptation. The focus of this paper is the exploration of temporal relation knowledge.

\section{Method}
\subsection{Problem formulation}
In a cross-user HAR problem, a labelled source user $\displaystyle S^{train} =\left\{\left( x_{i}^{train} ,\ y_{i}^{train}\right)\right\}_{i=1}^{n^{train}} $ drawn from a joint probability distribution $\displaystyle P^{Source}(X, y) $ and a target user $\displaystyle S^{test} =\left\{\left( x_{i}^{test} ,\ y_{i}^{test}\right)\right\}_{i=1}^{n^{test}} $ drawn from a joint probability distribution $\displaystyle P^{Target}(X, y) $, where $\displaystyle {n^{train}} $ and $\displaystyle {n^{test}} $ are the number of source and target samples respectively.  $\displaystyle S^{train}$ and $\displaystyle S^{test}$ have the same feature space and label space. In real-world applications, a source user and a target user are highly likely to have different data distributions, i.e., $\displaystyle P^{Source}(X, y) \neq P^{Target}(X, y) $. 
Given source user data $\displaystyle \left\{\left( x_{i}^{train} ,\ y_{i}^{train}\right)\right\}_{i=1}^{n^{train}} $ and target user data $\displaystyle \left\{\left( x_{i}^{test} \right)\right\}_{i=1}^{n^{test}} $, the goal is to achieve correct activity classification for unlabeled target user data.

\subsection{Motivation}
\begin{figure}[H]
\centering
\scalebox{1.0}{
  \includegraphics[width=\columnwidth]{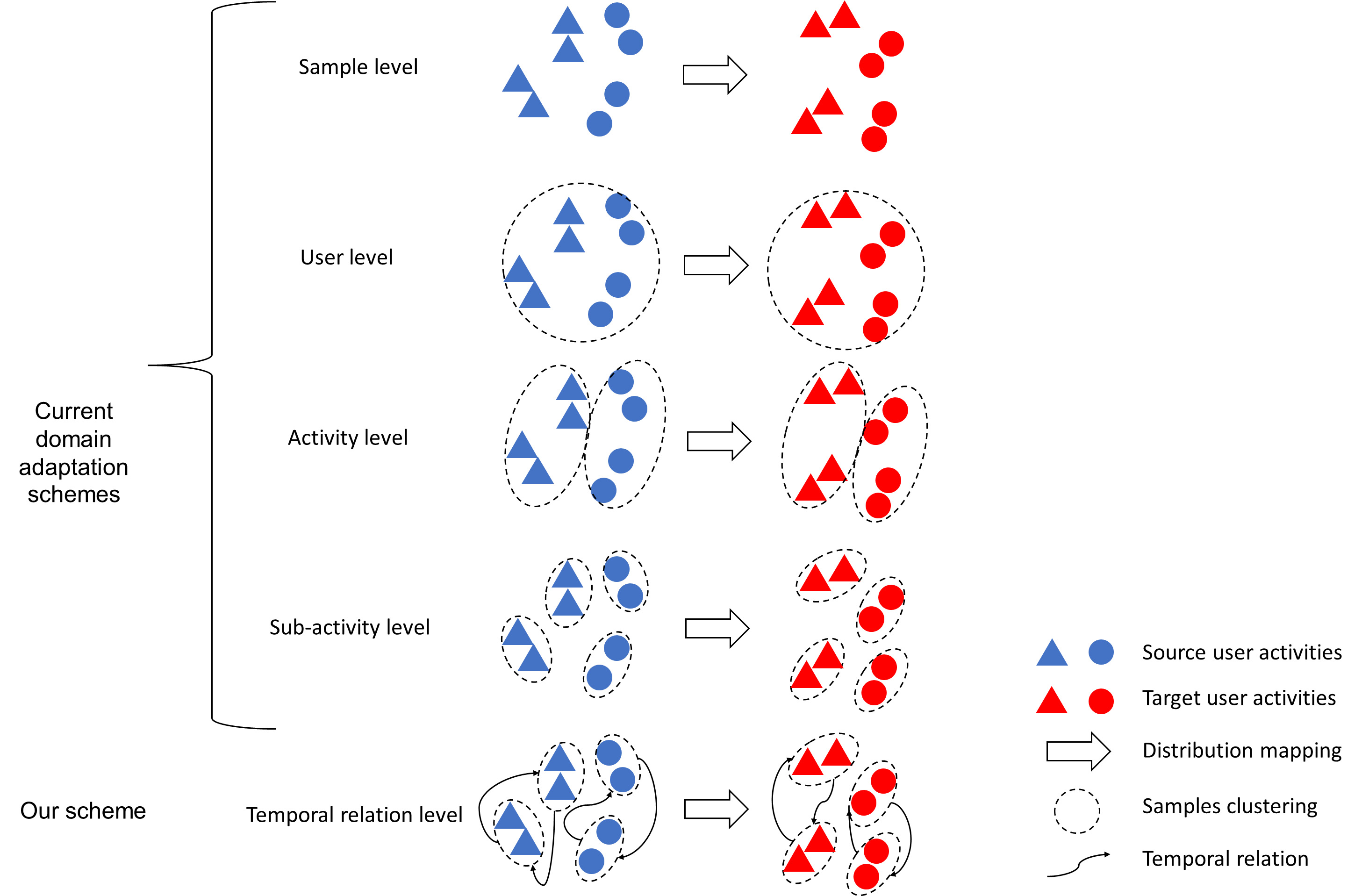}
}
\caption{Comparison of different domain adaptation schemes in cross-user HAR.\label{motivation}}
\end{figure}

In general, the principle and key idea of domain adaptation are to reduce the difference between the source domain(s) and the target domain(s) and bridge the gap in the data distributions between the domains. Existing cross-user HAR domain adaptation approaches are mainly based on sample level, user level, activity level and sub-activity level as shown in Figure~\ref{motivation}. However, these methods ignore the temporal relation across chronological time series samples and treat samples as $\displaystyle i.i.d. $ in different domains. In fact, temporal relation knowledge is an important property hidden in time series data that can be used as a bridge for cross-user domain adaptation. Different users' time series activity data follow the same temporal relation in various activities. For example, the walking activity includes three sub-activities, namely raising the leg, thrusting forward and feet to the ground, regardless of different users. These temporal relations are valuable knowledge that can better support data distribution alignment across different users. Therefore, we propose a new method called TROT that introduces temporal relation knowledge for solving domain adaptation.

\subsection{Temporal Relation Optimal Transport}

\begin{figure}[h!]
\centering
\scalebox{0.9}{
  \includegraphics[width=\columnwidth]{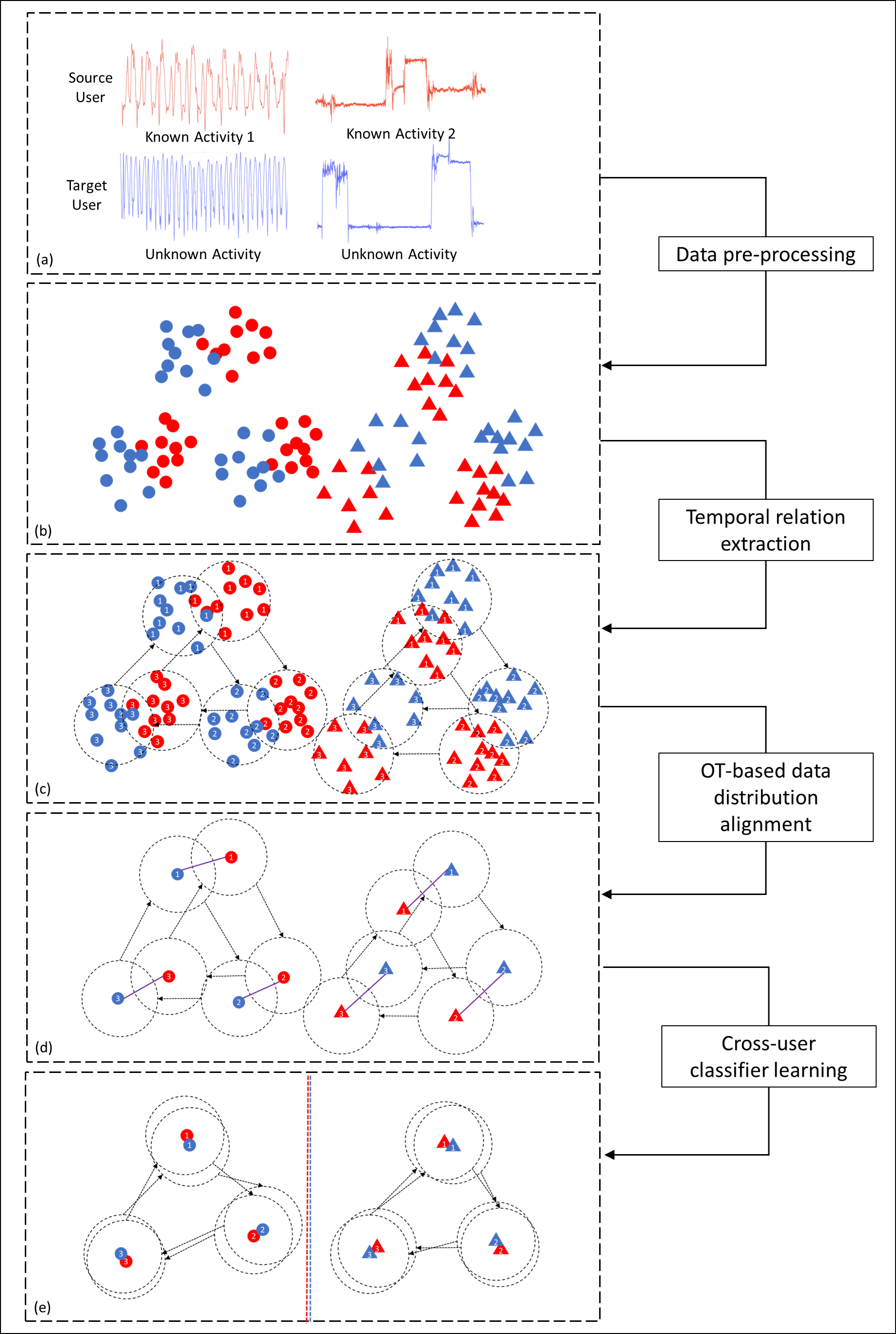}
}
\caption{TROT Learning Process Figure.\label{TROT_framework}}
\end{figure}

In this section, we propose an OT implementation that utilizes temporal relation knowledge called TROT. In TROT, HMM is utilized to capture temporal relation knowledge, and OT is used to align the data distributions between the source user and the target user. In addition, we propose to add a new regularizer which preserves temporal order information to enhance OT optimization for better domain adaptation performance. TROT method mainly includes four steps as listed below, and detailed explanations and discussions are provided in the following subsections. Figure~\ref{TROT_framework} shows the overall TROT learning process.

\begin{enumerate}
  \item \textbf{Data pre-processing}: Data segmentation and feature extraction for both source and target users is implemented as in Figure~\ref{TROT_framework} from figure (a) to figure (b). In Figure~\ref{TROT_framework}, the red colour is associated with the source user, while the blue colour is associated with the target user. The different shapes, such as circle and triangle, represents different activity classes. Sensor-based HAR data consists of observations taken sequentially over time. When working with time series data, the sliding window technique is often employed for data segmentation. Then, for each window of data, the time and frequency domain features are extracted. Time domain features capture raw temporal patterns in the data, while frequency domain features provide insight into the rhythmic or periodic components of the data. These features are crucial in cross-user HAR because they can capture general patterns of human movement that are less dependent on individual idiosyncrasies, thereby improving the model's ability to generalize across different users. In this step, the raw data are transformed into samples with features.
  \item \textbf{Temporal relation extraction}: Temporal state information is extracted via HMM for source and target users as in Figure~\ref{TROT_framework} from figure (b) to figure (c). The number in each shape indicates the corresponding temporal state information hidden in the time series data. For example, the numbers one, two and three in triangles may correspond to raising the leg, thrusting forward and feet to the ground of walking activity. The black dash arrow between the numbers means the temporal relation. The learned temporal relation knowledge is the common knowledge across users that is useful to bridge the gap between source and target users for better domain adaptation. In this step, the samples are clustered and form several clusters based on their temporal relation information.
  \item \textbf{OT-based data distribution alignment}: The data distribution alignment task is achieved by transforming the samples from source user to target user as in Figure~\ref{TROT_framework} from figure (c) to figure (d). The cluster centres are used to represent the corresponding cluster, and then the centres of the source user are projected to the centres of the target user considering the temporal order information for learning the mapping plan. The purple solid line in Figure~\ref{TROT_framework} means the data distribution transformation from the source user to the target user. In this step, the samples from the source user are transformed based on the learned mapping plan.
  \item \textbf{Cross-user classifier learning}: After the above data distribution alignment step between source and target users, the decision boundary of activity classes of the source and target users should be relatively consistent. Therefore, the transformed source user samples with labels are utilized for training a classifier for recognizing target user samples as in Figure~\ref{TROT_framework} from figure (d) to figure (e). The red dashed line is the decision boundary of activity classes of the source user, while the blue dashed line is the decision boundary of the target user. In this step, the learned classifier from the transformed source user is also suitable for the target user.
\end{enumerate}

The highlight of the TROT method is the temporal relation extraction step and the OT-based data distribution alignment step. Discussion is focused on these two steps.

\subsubsection{Temporal relation extraction}
 As in Figure~\ref{TROT_framework}, after the data pre-processing step, there is implicit temporal relation information that needs to be captured. Here, we consider HMM \cite{rabiner1986introduction} learning for temporal relation representation. HMM is the directed temporal probability graph model that describes the dependency of two related sequences. These two related sequences are called state sequence $\displaystyle S=(s_1,s_2,s_3,\ldots,s_T) $ and observation sequence $\displaystyle X=(x_1,x_2,x_3,\ldots,x_T) $. $\displaystyle Q=(q_1,q_2,q_3,\ldots,q_N) $ is the state set of all the possible states. The value of the state sequence at $\displaystyle t $ time point $\displaystyle s_t $ is only related to the value of the state sequence of $\displaystyle t-1 $ time point $\displaystyle s_{t-1} $, and the value of the observation sequence at $\displaystyle t $ time point $\displaystyle x_t $ is only related to the value of the state sequence at $\displaystyle t $ time point $\displaystyle s_t $. Here, there are three types of probability that HMM needs to learn: 1) $\displaystyle p(S_1) $ initial state probability, 2) $\displaystyle p(S_t|S_{t-1}) $ state transition probability and 3) $\displaystyle p(X_t|S_t) $ emission probability. 

For the cross-user HAR scenario, each sub-activity can be seen as a state. The projection of the observed samples and the corresponding sub-activity can be learned via HMM emission probability. The temporal relation between the sub-activities can be learned via HMM state transition probability. In this way, the hidden temporal information can be captured and represented. 

In detail, initial state probability is the possibility of the first sub-activity and is assumed to be the same across users in our study. Therefore, the first captured state is set as the initial set for the source user and the target user. Emission probability is the probability of a sample occurring, given a specified sub-activities (i.e. state) at time point $\displaystyle t $. The observation sequence, $\displaystyle X_t $, can be expressed as a function of $\displaystyle S_t $ based on the learning of emission probability, which means $\displaystyle X_t=f(S_t) $. For the emission probability, we assume it conforms to Gaussian distribution. The emission probability in each class can be expressed as $\displaystyle B=[ b_{k}]_{N\times 1} ,\ b_{k} =\mathcal{N}( \mu _{k} ,\ \sigma _{k}) ,\ X_{k} \sim b_{k} ,\ k=1,2,...,N$, $\displaystyle N $ is the number of sub-activities in each class. Here, $\displaystyle \mu_{k} $ means the centre value of $\displaystyle k^{th} $ state’s Gaussian distribution, $\displaystyle \sigma _{k} $ means the covariance value of $\displaystyle k^{th} $ state’s Gaussian distribution, and $\displaystyle X_{k} $ means the samples belong to the $\displaystyle k^{th} $ state. Assuming there are C activity classes, then there are a total of $\displaystyle N \times  C $ Gaussian distributions need to be learned for each user, i.e., $\displaystyle B^{Source}=[\mathbf{b}^{Source}]_{N\times C}$ and $\displaystyle B^{Target}=[\mathbf{b}^{Target}]_{N\times C} $. In this way, we abstract the observation samples to a higher level of sub-activities, and then the temporal relation extraction happens at the sub-activities level.

The state transition probability is the probability of a sub-activities at time point $\displaystyle t-1 $ change to another sub-activity at time point $\displaystyle t $. For the state transition probability, we reduce the difference between the source user and the target user via common temporal relation knowledge of the human activity. The common knowledge is that the state transition follows the sub-activities order because of the continuity of human activities, which means the current state can only change to the following state in a future time point. For example, walking activity should first raise the leg and then thrust forward in the next state, but not jump to the last state of feet to the ground step. Therefore, $\displaystyle p(S_t|S_{t-1}) $ state transition probability can be expressed as a $\displaystyle N \times N $ matrix $\displaystyle A=[a_{i,j}]_{N\times N} ,\ a_{i,j} =P( S_{t} =q_{j} |S_{t-1} =q_{i}) ,\ i,j=0,1,2,...,N, i \neq j$. $\displaystyle a_{i,j} $ means the transition probability from state $\displaystyle q_i $ in $\displaystyle t-1 $ time point to state $\displaystyle q_j $ in $\displaystyle t $ time point. There are total $\displaystyle N $ sub-activities. $\displaystyle q_0 $ is the initial state and $\displaystyle q_N $ is the ending state. If an activity such as running or going upstairs has periodic cycle characteristics, then $\displaystyle q_0=q_N $. The form of the state transition matrix for HMM is given below. Here, we set $\displaystyle \forall i\ \ a_{i,i+1}=1 $ to make the state transition probability a common temporal relation knowledge in both the source and target users.

\begin{center}
$\begin{bmatrix}
a_{0,1} & 0 & \cdots & \cdots & 0 & 0 \\
0 & a_{1,2} & 0 & \cdots & \cdots & 0 \\
0 & 0 & a_{2,3} & 0 & \cdots & \vdots \\
\vdots & \cdots & \cdots & \cdots & \cdots & 0 \\
0 & \cdots & \cdots & 0 & 0 & a_{N-1,N} \

\end{bmatrix}$
\end{center}

 With the observation sequence $\displaystyle X $, our goal is to calculate the parameters of emission probability in each activity for each user. The process of parameter learning can be achieved by Expectation Maximum (EM) \cite{dempster1977maximum} algorithm. After the HMM parameter learning, the source user data distribution can be expressed as $\displaystyle B^{Source} $, and the target user can be expressed as $\displaystyle B^{Target} $. As Figure~\ref{TROT_framework} (c) shows, now each class is represented by multiple temporally-related Gaussian distributions. Our next task is the alignment of sub-activities in each class between source and target users.

\subsubsection{Optimal Transport based data distribution alignment}

Once the temporal-related distribution is extracted by HMM, the data distribution alignment is achieved via an OT-based method that maps the Gaussian distributions between the source and target users. OT has drawn attentions in recent years for domain adaptation, and it aims at moving source domain data distribution on top of target domain data distribution with an optimization transportation plan \cite{kerdoncuff2021metric}.

Based on the OT-based method, the source user and target user can be expressed as data distributions respectively: 
\begin{align*}
    P^{Source} &= \sum_{i=1}^{i=k^{Source}} w_{i}^{Source} \delta_{\mu_{i}^{Source}}, \\
    P^{Target} &= \sum_{j=1}^{j=k^{Target}} w_{j}^{Target} \delta_{\mu_{j}^{Target}}.
\end{align*}
Here, $\sum_{i=1}^{i=k^{Source}} w_{i}^{Source} = 1$, and $\sum_{j=1}^{j=k^{Target}} w_{j}^{Target} = 1$. $\boldsymbol{\mu}$ is the center value of Gaussian distributions in $B$ learned from HMM, and $\boldsymbol{\delta_{\mu}}$ is the Dirac function corresponding to the location of $\boldsymbol{\mu}$. The probabilities $\boldsymbol{w}$ are the probability masses associated with $\boldsymbol{\mu}$. In our case, $k^{Source} = k^{Target} = C \times N$, where $C$ is the number of activity classes, and $N$ is the number of states. We set $\forall i, j\ w_{i}^{Source} = w_{j}^{Target} = \frac{1}{C \times N}$.

Similar to \cite{flamary2016optimal}, we also apply regularization approach to OT to prevent overfitting and keep structural risk minimization. Entropy regularization \cite{cuturi2013sinkhorn} $\displaystyle H( \gamma )$ is introduced with the idea of lowering the sparsity of coupling matrix $\displaystyle \gamma $ via increasing its entropy to find a smoother optimal transport plan. Here, the coupling matrix represents the transport plan between two probability distributions as source user and target user. The smoother optimal transport plan often translates to being more generalizable, which is beneficial for domain adaptation, instead of having very specific, point-to-point transport, which might overfit the source user. Group-sparse regularization \cite{flamary2016optimal} $\displaystyle \si{\ohm}( \gamma )$ considers the use of label distribution knowledge that makes each sample in the target domain receive masses only from the samples in the source domain with the same label. By preserving the inherent label structure of the data, group-sparse regularization promotes the correct transfer of class structures between domains and makes models more generalizable to the target user.

\begin{figure}[H]
\centering
\includegraphics[width=5cm]{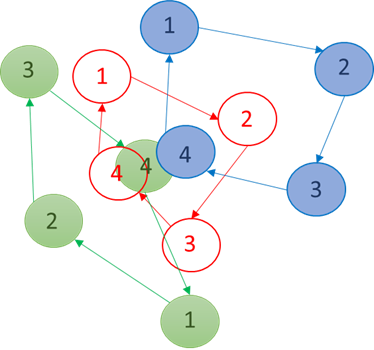}
\caption{OT-based Temporal Order Preservation Regularization Figure.\label{TRDA_regularization}}
\end{figure}

Besides, we propose to add a new regularization term to the regularized optimal transport that preserves temporal order information. The main intuition underlying this temporal order preservation regularization is that even two Gaussian distribution centres may be closer between the source user and target user; however, different temporal relation order may lead to wrong data distribution alignment. As Figure~\ref{TRDA_regularization} shows, the red is an activity from the source user, and the blue and green are two different activities from the target user. In this example, both green and blue are very close to red in terms of the overall distance between the centres of sub-activities distributions. However, blue is the only suitable one after considering the temporal order of the state Gaussian distribution.

\begin{table}[H]
\caption{Temporal Order Mapping Between Source and Target Users Table.\label{Temporal_Order_Mapping}}
\centering
\resizebox{0.3\columnwidth}{!}{%
\begin{tabular}{llllllllll}
 & \multicolumn{1}{l|}{} & \multicolumn{1}{l|}{1} & \multicolumn{1}{l|}{2} & \multicolumn{1}{l|}{3} & \multicolumn{1}{l|}{4} & \multicolumn{1}{l|}{5} & \multicolumn{1}{l|}{6} & \multicolumn{1}{l|}{7} & \multicolumn{1}{l|}{8} \\ \cline{3-10} 
 & \multicolumn{1}{l|}{} & \multicolumn{1}{l|}{\cellcolor[HTML]{67FD9A}\textbf{1}} & \multicolumn{1}{l|}{\cellcolor[HTML]{67FD9A}\textbf{2}} & \multicolumn{1}{l|}{\cellcolor[HTML]{67FD9A}\textbf{3}} & \multicolumn{1}{l|}{\cellcolor[HTML]{67FD9A}\textbf{4}} & \multicolumn{1}{l|}{\cellcolor[HTML]{8DA7F4}\textbf{1}} & \multicolumn{1}{l|}{\cellcolor[HTML]{8DA7F4}\textbf{2}} & \multicolumn{1}{l|}{\cellcolor[HTML]{8DA7F4}\textbf{3}} & \multicolumn{1}{l|}{\cellcolor[HTML]{8DA7F4}\textbf{4}} \\ \hline
\multicolumn{1}{l|}{1} & \cellcolor[HTML]{FD6864}\textbf{1} & \checkmark &  &  &  & \checkmark &  &  &  \\ \hline
\multicolumn{1}{l|}{2} & \cellcolor[HTML]{FD6864}\textbf{2} &  & \checkmark &  &  &  & \checkmark &  &  \\ \hline
\multicolumn{1}{l|}{3} & \cellcolor[HTML]{FD6864}\textbf{3} &  &  & \checkmark &  &  &  & \checkmark &  \\ \hline
\multicolumn{1}{l|}{4} & \cellcolor[HTML]{FD6864}\textbf{4} &  &  &  & \checkmark &  &  &  & \checkmark \\ \hline
\multicolumn{1}{l|}{5} & \cellcolor[HTML]{FFFE65}\textbf{1} & \checkmark &  &  &  & \checkmark &  &  & \\ \hline
\multicolumn{1}{l|}{6} & \cellcolor[HTML]{FFFE65}\textbf{2} &  & \checkmark &  &  &  & \checkmark &  &  \\ \hline
\multicolumn{1}{l|}{7} & \cellcolor[HTML]{FFFE65}\textbf{3} &  &  & \checkmark &  &  &  & \checkmark &  \\ \hline
\multicolumn{1}{l|}{8} & \cellcolor[HTML]{FFFE65}\textbf{4} &  &  &  & \checkmark &  &  &  & \checkmark \\ \hline
\end{tabular}%
}
\end{table}

This temporal order preservation regularizer is defined as: 
\begin{equation*}
\begin{aligned}
T( \gamma ) ={\textstyle \sum _{i}\begin{Vmatrix}
\gamma ( i,I_{i})
\end{Vmatrix}_{2}} 
\end{aligned}
\end{equation*}

The coupling matrix $\displaystyle \gamma $ has the indices of the column, each of which corresponds to a Gaussian distribution from the target user, while the indices of the row correspond to a Gaussian distribution from the source user. For example, in Table~\ref{Temporal_Order_Mapping}, the indices of the coupling matrix are from 1 to 8 for both the column and the row. $\displaystyle I_{i} $ is a set of indices that tell which columns in $\displaystyle \gamma $ correspond to the same temporal order as the $\displaystyle i^{th} $ Gaussian distribution of the source user. In other words, $\displaystyle I_{i} $ helps to find the matching temporal order in the target user for each Gaussian distribution in the source user. As in Table~\ref{Temporal_Order_Mapping},  the red and yellow are two activities from the source user, while the blue and green are two activities from the target user. All the activities have four temporal states (i.e. Gaussian distributions). The number from 1 to 4 in each activity represents the temporal order. For example, the temporal order 1 in red activity of the source user matches the temporal order 1 in blue and green activities of the target user, as the checkmark symbol shows. $\displaystyle \gamma(i,I_{i}) $ is a vector that is achieved by picking out the coefficients from the $\displaystyle i^{th}$ row of $\displaystyle \gamma $ that correspond to these matched temporal orders from the column. These coefficients are the transport plan from the $\displaystyle i^{th} $ Gaussian distribution of the source user to the matched Gaussian distributions in the target user. 

In summary, this regularizer encourages the Gaussian distributions of the source user to match the Gaussian distributions with the same temporal order of the target user. For example, in Figure~\ref{TRDA_regularization}, order 1 of red is encouraged to map to order 1 of blue instead of mapping to order 3 of green.

With this temporal order preservation regularization, the masses corresponding to each Gaussian distribution of specified temporal order value match samples of the source and target users with the same temporal order value. So, the overall goal of TROT is:

\begin{equation*}
\begin{aligned}
\gamma^* &= \underset{\gamma}{\arg\min} \langle \gamma, \mathcal{C} \rangle _F +\lambda H( \gamma ) +\eta \si{\ohm}( \gamma ) +\tau T( \gamma ) 
\end{aligned}
\end{equation*} 
\begin{equation*}
\text{s.t.} 
\end{equation*} 
\begin{equation*}
\begin{aligned}
\gamma \mathbf{1}_{k^{\text{Target}}} = P^{\text{Source}} 
\end{aligned}
\end{equation*} 
\begin{equation*}
\begin{aligned}
\gamma^T \mathbf{1}_{k^{\text{Source}}} = P^{\text{Target}}
\end{aligned}
\end{equation*}

$\displaystyle \gamma \in \mathbb{R}^{k^{\text{Source}} \times k^{\text{Target}}} $ is the coupling matrix between the probability distributions of the source domain and target domain, $\displaystyle \mathbf{1}_k $ is k-dimensional vectors of ones, $\displaystyle \langle .,. \rangle _F $ is the Frobenius dot product. $\displaystyle \mathcal{C} $ is the cost matrix of the squared Euclidean distance between two locations, $\displaystyle \mathcal{C}(i,j)=\begin{Vmatrix}
\mu _{i}^{Source} -\mu _{j}^{Target}
\end{Vmatrix}_{2}^{2} $, denotes the cost of pushing forward a probability mass from $\displaystyle \mu _{i}^{Source} $ to $\displaystyle \mu _{j}^{Target} $. $\displaystyle \lambda, \eta, \tau$ are coefficients.

The generalized conditional gradient (GCG) approach is used to solve the regularized optimization problem \cite{flamary2016optimal}. GCG algorithm addresses the general case of constrained minimization of composite functions defined as 

\begin{equation*}
\begin{aligned}
 \underset{\gamma}{\min}f(\gamma)+g(\gamma)
\end{aligned}
\end{equation*}

$\displaystyle f(\gamma) $ is a differentiable and possibly non-convex function, $\displaystyle g(\gamma) $ is a convex, possibly non-differentiable function. Here, we set 

\begin{equation*}
\begin{aligned}
 f(\gamma)= \langle \gamma, \mathcal{C} \rangle _F +\lambda H( \gamma )
\\
 g(\gamma)=\eta \si{\ohm}( \gamma ) +\tau T( \gamma ) \\
 \text{s.t.} \quad  H'(\gamma)
\end{aligned}
\end{equation*}

Now, the TROT task is 
\begin{equation*}
\begin{aligned}
 \gamma^* = \underset{\gamma}{\arg\min} \langle \gamma, \mathcal{C}+\lambda \nabla H( \gamma^{iter} )\rangle _F+(\eta \si{\ohm}( \gamma ) +\tau T( \gamma ))
\end{aligned}
\end{equation*}

$\displaystyle iter $ is the number of iterations. After getting the optimal coupling matrix $\displaystyle \gamma^* $, barycentric mapping of the source domain is implemented for mapping the source domain to the target domain as 

\begin{equation*}
\begin{aligned}
\widehat{P^{Target}}=diag{(\gamma^\ast\mathbf{1}_{k^{Target}})}^{-1}\gamma^\ast P^{Target} 
\end{aligned}
\end{equation*}

In the final step, transformed source user data is utilized to train a classifier for classifying target user samples. Here, any classifier can be applied to learn a model with $\displaystyle\widehat{P^{Target}} $ and the labels of the source domain $\displaystyle y^{Source} $. We use the 1-Nearest Neighbour classification method to get labels $\displaystyle \{y_j^{Target}\}_{j=1}^{n^{Target}} $. 1-Nearest Neighbour is a simple instance of the k-Nearest Neighbours algorithm \cite{guo2003knn}, which is a type of instance-based learning. Given a sample from the target user, the algorithm searches the entire samples from the source user to find the single sample that is closest to the sample from the target user. Here, the distance metric is the Euclidean metric. The label of the nearest sample from the source user is then assigned to the sample from the target user.

\section{Experiments}

\subsection{Datasets and pre-processing}

We used three common sensor-based HAR public datasets as Table~\ref{tab_datasets_info} shows for validating our cross-user transfer learning method. Here, we only use the sensor values of the accelerometer and gyroscope in the position of the right lower arm to explore the practical scenario of using the smartwatch. In the following, the information of each dataset is introduced briefly.

\begin{table}[H]
\caption{Three sensor-based HAR datasets information Table.\label{tab_datasets_info}}
\centering
\resizebox{\columnwidth}{!}{%
\begin{tabular}{|l|l|l|l|}
\textbf{Dataset \& Ref.} & \textbf{Subjects} & \textbf{\#Activities} & \textbf{Common Activities} \\ \hline
OPPT & S1, S2, S3 & 4 & lying, sitting,   standing, walking \\ \hline
PAMAP2 & 1, 5, 6 & 11 & \begin{tabular}[c]{@{}l@{}}lying, sitting, standing, walking, \\ running, cycling, Nordic walking, \\ ascending stairs, descending stairs, \\ vacuum cleaning, ironing\end{tabular} \\ \hline
DSADS & 2, 4, 7 & 19 & \begin{tabular}[c]{@{}l@{}}sitting, standing, lying on back, \\ lying on right, ascending stairs, \\ descending stairs, standing in an elevator still, \\ moving around in an elevator, \\ walking in a parking lot, \\ walking on a treadmill in flat, \\ walking on a treadmill inclined positions, \\ running on a treadmill in flat, \\ exercising on a stepper, \\ exercising on a cross trainer, \\ cycling on an exercise bike in horizontal positions, \\ cycling on an exercise bike in vertical positions, \\ rowing, jumping, playing basketball\end{tabular}
\end{tabular}%
}
\end{table}

OPPORTUNITY (OPPT) \cite{chavarriaga2013opportunity} dataset contains recordings of subjects in a daily living scenario performing morning activities. During the recordings, subjects perform the activities without any restriction by following a loose description of the overall actions to perform. The sampling frequency is 30 Hz. Physical activity monitoring (PAMAP2) dataset \cite{reiss2012introducing} is collected following a protocol of specified activities for each subject. Over 10 hours of data were collected altogether. The sampling frequency is 100 Hz. Daily and Sports Activities Data Set (DSADS) \cite{barshan2014recognizing} is collected by asking the subjects to perform the activities in their own fashion. Without any given instructions, this is likely to lead to greater inter-subject variations, and would be more similar to real-life situations. Each activity is performed by each subject for 5 min. Sensor units are calibrated to acquire data at 25 Hz sampling frequency. The three datasets are selected in order of increasing domain adaptation task difficulty (DSADS 19 activities $\displaystyle > $ PAMAP2 11 activities$\displaystyle > $ OPPT 4 activities). In this case, DSADS dataset contains more activity classes and more similar activities (3 different walking activities), which make it harder to distinguish.  

The data pre-processing step for all the methods follows the same setup as the TROT data segmentation and feature extraction step. Time series data is continuous sequence data. The sliding window technique is a common data segmentation method. For the purpose of temporal relation extraction, each window is set as a fixed time interval. Each window is set as 3s, and overlapping is 50\% as the common setting in sensor-based HAR task \cite{wang2018impact}. Following a similar feature extraction setup as \cite{wang2018stratified}, the $\displaystyle x,y,z$ axis data in each sensor of accelerometer and gyroscope are combined as $\displaystyle \sqrt{x^{2} +y^{2} +z^{2}}$, 19 features from both time and frequency domains are extracted for a single sensor as in Table~\ref{tab:features_extraction_table}. Therefore, in each window, there are total of 38 features because of two sensors of accelerometer and gyroscope. Besides, standardization of maximum absolute value scaler is applied to scales and translates each feature individually in the range of $\displaystyle [-1,1] $ to prevent the side effect of the features with different numerical order.

\begin{table}[H]
\caption{Features Extraction Per Sensor}
\label{tab:features_extraction_table}
\centering
\resizebox{0.75\columnwidth}{!}{%
\begin{tabular}{lll}
\hline
\textbf{ID} & \textbf{Feature} & \textbf{Description} \\ \hline
1 & Mean & Average value of samples in window \\
2 & Var & Variance \\
3 & STD & Standard deviation \\
4 & Mode & The value with the largest frequency \\
5 & Maximum & Maximum \\
6 & Minimum & Minimum \\
7 & Mean crossing rate & Rate of times signal crossing mean value \\
8 & Range & Maximum minus minimum \\
9 & DC & Direct component \\
10-14 & Five amplitude features & Mean, variance, STD, skewness, kurtosis \\
15-19 & FFT variance & Mean, variance, STD, skewness, kurtosis \\ \hline
\end{tabular}%
}
\end{table}

\subsection{Experiments and methods comparison}
Here are the categories of methods for comparison, including no domain adaptation methods and domain adaptation methods.

\begin{enumerate}
    \item No domain adaptation: 
    \begin{enumerate}
        \item NA: No Adaptation
        \item TD: Target domain 1NN
    \end{enumerate}
    \item Domain adaptation:
    \begin{enumerate}
        \item The usage of temporal relation: TROT
        \item No usage of temporal relation:
        \begin{enumerate}
            \item LMNN: Large Margin Nearest Neighbour \cite{weinberger2009distance}
            \item SA: Subspace Alignment \cite{fernando2013unsupervised}
            \item OT: Optimal transport \cite{cuturi2013sinkhorn}
            \item CORAL: CORrelation Alignment \cite{sun2016return}
            \item OTDA: Optimal transport for domain adaptation \cite{flamary2016optimal}
            \item SOT: Substructural optimal transport \cite{lu2021cross}
        \end{enumerate}
    \end{enumerate}
\end{enumerate}

The tables provide a comprehensive comparison of different methods applied to the OPPT, PAMAP2, and DSADS datasets. The performance of each method is evaluated based on the transition from one user to another user. Three users are randomly selected from all the users in each dataset. Then, a one-to-one cross-user HAR task is implemented, encompassing all possible pairings between users in each dataset.

For the NA method, the 1-Nearest Neighbour model is trained by source user samples and is applied directly to target user samples without any domain adaptation mapping. For the TD method, the 1-Nearest Neighbour model is trained and predicted by target user samples, which is the optimal baseline that we aim to approach. For the domain adaptation methods, hyper-parameters tuning is performed, and we follow a similar process as suggested in \cite{flamary2016optimal} to prevent overfitting on the testing set. The target user is partitioned into two equal datasets the validation set and the test set. For the TROT method that captures the temporal relation, it is particularly crucial to assess its performance on unseen future time-series data that exhibit a distinction from the data utilized for aligning the distribution. Therefore, the target domain is partitioned in equal time duration of two parts that keep temporal order. The validation set is used to obtain the combination of hyper-parameters that achieve the highest accuracy from the hyper-parameter space. Then, the performance of the testing set is evaluated with the selected combination of hyper-parameters. Classification accuracy on the target domain is the evaluation metric that we used.

\begin{table}[H]
\caption{OPPT Dataset Methods Comparison Table}
\label{tab:oppt_result_table}
\centering
\resizebox{0.9\textwidth}{!}{%
\begin{tabular}{|l|c|c|c|c|c|c|}
\hline
\textbf{Method} & \textbf{S1 \(\rightarrow\) S2} & \textbf{S1 \(\rightarrow\) S3} & \textbf{S2 \(\rightarrow\) S1} & \textbf{S2 \(\rightarrow\) S3} & \textbf{S3 \(\rightarrow\) S1} & \textbf{S3 \(\rightarrow\) S2} \\ \hline
NA & 44.18 & 54.11 & 50.36 & 53.22 & 47.94 & 46.53 \\ \hline
TD & 100 & 100 & 100 & 100 & 100 & 100 \\ \hline
LMNN & 77.77 & 73.10 & 75.04 & 69.94 & 65.96 & 71.06 \\ \hline
SA & 79.50 & 84.11 & 83.39 & 83.22 & 78.18 & 76.53 \\ \hline
OT & 80.37 & 81.33 & 81.69 & 79.05 & 81.69 & 78.89 \\ \hline
CORAL & 74.18 & 84.74 & 80.36 & 84.74 & 78.67 & 75.92 \\ \hline
OTDA & 83.96 & 80.44 & 86.42 & 78.04 & 86.66 & 80.62 \\ \hline
SOT & 75.60 & 79.94 & 82.76 & 73.62 & 74.09 & 69.53 \\ \hline
TROT & \textbf{100} & \textbf{100} & \textbf{100} & \textbf{100} & \textbf{100} & \textbf{100} \\ \hline
\end{tabular}%
}
\end{table}

In general, all the domain adaptation methods can achieve increased performance than the baseline method of NA in the OPPT dataset as shown in Table~\ref{tab:oppt_result_table}. TROT method achieved the highest accuracy across all the six tasks, with a perfect score of 100\%. This suggests that the method is extremely efficient and reliable for the OPPT dataset. OTDA also showed promising results, especially in the tasks from S2 to S1 and from S3 to S1. SA and OT methods were relatively consistent in their performance across different tasks. The SA method, in particular, performed quite well, especially in the task from S1 to S3 and S2 to S1.

\begin{table}[H]
\centering
\caption{PAMAP2 Dataset Methods Comparison Table}
\label{tab:pamap2_result_table}
\resizebox{0.8\textwidth}{!}{%
\begin{tabular}{|l|c|c|c|c|c|c|}
\hline
\textbf{Method} & \textbf{1 \(\rightarrow\) 5} & \textbf{1 \(\rightarrow\) 6} & \textbf{5 \(\rightarrow\) 1} & \textbf{5 \(\rightarrow\) 6} & \textbf{6 \(\rightarrow\) 1} & \textbf{6 \(\rightarrow\) 5} \\ \hline
NA & 28.72 & 42.02 & 47.91 & 55.46 & 60.43 & 35.75 \\ \hline
TD & 100 & 100 & 100 & 100 & 100 & 100 \\ \hline
LMNN & 22.25 & 20.17 & 26.04 & 26.41 & 36.03 & 21.34 \\ \hline
SA & 31.33 & 43.58 & 49.81 & 58.82 & 60.18 & 40.07 \\ \hline
OT & 53.69 & 61.70 & 57.77 & 61.22 & 57.14 & 57.32 \\ \hline
CORAL & 32.69 & 42.62 & 50.95 & 57.14 & 59.04 & 38.25 \\ \hline
OTDA & 64.47 & 65.23 & 63.08 & 65.19 & 66.12 & 64.47 \\ \hline
SOT & 69.82 & 57.80 & 64.54 & 61.95 & 62.48 & 56.14 \\ \hline
TROT & \textbf{84.80} & \textbf{75.67} & \textbf{78.28} & \textbf{83.05} & \textbf{82.46} & \textbf{73.76} \\ \hline
\end{tabular}%
}
\end{table}

When applied to the PAMAP2 dataset in Table~\ref{tab:pamap2_result_table}, all the domain adaptation methods except for LMNN achieved better performance than NA in most of the cross-user tasks.  Negative transfer learning (i.e., worse performance) happened on LMNN for all the tasks of the PAMAP2 dataset. TROT maintained the best performance compared to other methods. Moreover, SOT and OTDA showed notable performance, with SOT showing particularly high efficiency in the tasks from 1 to 5 and 5 to 1. OTDA showed a good performance in the tasks from 1 to 6 and from 6 to 1.

\begin{table}[H]
\caption{DSADS Dataset Methods Comparison Table}
\label{tab:dsads_result_table}
\centering
\resizebox{0.8\textwidth}{!}{%
\begin{tabular}{|l|c|c|c|c|c|c|}
\hline
\textbf{Method} & \textbf{2 \(\rightarrow\) 4} & \textbf{2 \(\rightarrow\) 7} & \textbf{4 \(\rightarrow\) 2} & \textbf{4 \(\rightarrow\) 7} & \textbf{7 \(\rightarrow\) 2} & \textbf{7 \(\rightarrow\) 4} \\ \hline
NA & 48.57 & 41.12 & 39.76 & 38.61 & 44.24 & 37.88 \\ \hline
TD & 100 & 100 & 100 & 100 & 100 & 100 \\ \hline
LMNN & 18.45 & 16.47 & 13.13 & 15.63 & 11.78 & 17.67 \\ \hline
SA & 48.83 & 40.91 & 39.86 & 38.30 & 43.04 & 38.82 \\ \hline
OT & 47.26 & 35.64 & 46.59 & 38.09 & 36.89 & 36.63 \\ \hline
CORAL & 49.35 & 41.27 & 40.70 & 39.24 & 43.30 & 39.76 \\ \hline
OTDA & 48.10 & 38.25 & 45.23 & 45.28 & 35.28 & 43.30 \\ \hline
SOT & 43.23 & 46.72 & 48.47 & 45.64 & 43.00 & 41.11 \\ \hline
TROT & \textbf{53.95} & \textbf{52.47} & \textbf{49.27} & \textbf{47.21} & \textbf{53.68} & \textbf{46.37} \\ \hline
\end{tabular}%
}
\end{table}

On the DSADS dataset in Table~\ref{tab:dsads_result_table}, CORAL, SOT, OTDA and TROT achieved higher performance than NA in more than half of the tasks in the DSADS dataset. TROT maintained its superior performance. Among other methods, TROT performed best, especially in the task from 2 to 4. SOT showed reasonable performance, especially in the tasks from 2 to 7 and from 4 to 2. OTDA also performed relatively well but had a decrease in performance compared to the PAMAP2 dataset. LMNN had the lowest performance among all the methods, much like in the other two datasets. This suggests that LMNN might not be the best choice for cross-user HAR task. 

TROT method gets the highest accuracy among all the three datasets of 18 cross-user HAR tasks, the performance of other methods varies based on the dataset and the specific tasks. The results showed that the methods without considering temporal relation knowledge have difficulties (i.e. unstable performance) when performing time series transfer learning tasks, because the design rationale assumes the samples follow $\displaystyle i.i.d. $ in each domain. In contrast, TROT relaxes the constraint condition of domain adaptation without the $\displaystyle i.i.d. $ assumption of samples. Besides, TROT can capture temporal relation knowledge in each domain and reduce the difference by taking into account the temporal relation between the source and target users, which showed superior performance in the time series domain adaptation problem.

\section{Conclusion}

Current works in domain adaptation are mainly based on the assumption that samples in each domain are $\displaystyle i.i.d. $ drawn from a joint probability distribution. However, existing research works overlook the temporal relation among the samples in the process of aligning data distribution and considering the temporal relation knowledge embedded in time series data. In this paper, we propose the TROT method to utilize temporal relation and relax the $\displaystyle i.i.d. $ assumption for the source and target users and an OT and HMM based implementation for cross-user HAR. A new regularization term that preserves temporal order information for the regularized optimal transport mapping is proposed to extend the OT capability. Compared to existing domain adaptation, TROT obtains the temporal relation representation and completes the feature-based data distribution alignment to solve domain adaptation problem. Comprehensive experiments applied to three HAR public datasets demonstrate the superiority of the TROT method over other state-of-the-art methods.

In the future, we plan to remove the Markov assumption and extend HAR to capture and align more complex temporal relations in sensor-based HAR cross-user domain adaptation. Moreover, in the pursuit of refining and enhancing the applicability of the TROT, it is imperative to delineate the specific scenarios under which this approach may exhibit suboptimal performance. A critical aspect warranting investigation is the conceptualization and quantification of the "distance" between users or their corresponding data. This "distance" metric could potentially encapsulate the dissimilarity in types, positions, and other attributes of sensors utilized by different users. Establishing a threshold for this "distance" could provide a criterion to ascertain whether the application of TROT would yield satisfactory results. Such an exploration could illuminate the constraints of TROT, and guide the optimization of its algorithm to cater to a broader spectrum of user scenarios. Through a comprehensive analysis of these facets, the endeavor is to elevate the robustness and versatility of TROT, thus propelling it closer to being a universally applicable solution.

\bibliographystyle{splncs04} 
\bibliography{ref}

\begin{thebibliography}{10}
\providecommand{\url}[1]{\texttt{#1}}
\providecommand{\urlprefix}{URL }
\providecommand{\doi}[1]{https://doi.org/#1}

\bibitem{abdallah2018activity}
Abdallah, Z.S., Gaber, M.M., Srinivasan, B., Krishnaswamy, S.: Activity recognition with evolving data streams: A review. ACM Computing Surveys (CSUR)  \textbf{51}(4),  1--36 (2018)

\bibitem{amft2005detection}
Amft, O., Junker, H., Troster, G.: Detection of eating and drinking arm gestures using inertial body-worn sensors. In: Ninth IEEE international symposium on wearable computers (ISWC'05). pp. 160--163. IEEE (2005)

\bibitem{andreas2014tutorial}
Andreas, B., Blanke, U., Schiele, B.: A tutorial on human activity recognition using body-worn inertial sensors. ACM Computing Surveys (CSUR)  \textbf{46}(3), ~33 (2014)

\bibitem{barshan2014recognizing}
Barshan, B., Y{\"u}ksek, M.C.: Recognizing daily and sports activities in two open source machine learning environments using body-worn sensor units. The Computer Journal  \textbf{57}(11),  1649--1667 (2014)

\bibitem{bulling2012multimodal}
Bulling, A., Ward, J.A., Gellersen, H.: Multimodal recognition of reading activity in transit using body-worn sensors. ACM Transactions on Applied Perception (TAP)  \textbf{9}(1),  1--21 (2012)

\bibitem{chavarriaga2013opportunity}
Chavarriaga, R., Sagha, H., Calatroni, A., Digumarti, S.T., Tr{\"o}ster, G., Mill{\'a}n, J.d.R., Roggen, D.: The opportunity challenge: A benchmark database for on-body sensor-based activity recognition. Pattern Recognition Letters  \textbf{34}(15),  2033--2042 (2013)

\bibitem{chen2021deep}
Chen, K., Zhang, D., Yao, L., Guo, B., Yu, Z., Liu, Y.: Deep learning for sensor-based human activity recognition: Overview, challenges, and opportunities. ACM Computing Surveys (CSUR)  \textbf{54}(4),  1--40 (2021)

\bibitem{chen2019cross}
Chen, Y., Wang, J., Huang, M., Yu, H.: Cross-position activity recognition with stratified transfer learning. Pervasive and Mobile Computing  \textbf{57},  1--13 (2019)

\bibitem{cuturi2013sinkhorn}
Cuturi, M.: Sinkhorn distances: Lightspeed computation of optimal transport. Advances in neural information processing systems  \textbf{26} (2013)

\bibitem{dempster1977maximum}
Dempster, A.P., Laird, N.M., Rubin, D.B.: Maximum likelihood from incomplete data via the em algorithm. Journal of the royal statistical society: series B (methodological)  \textbf{39}(1),  1--22 (1977)

\bibitem{deng2014cross}
Deng, W.Y., Zheng, Q.H., Wang, Z.M.: Cross-person activity recognition using reduced kernel extreme learning machine. Neural Networks  \textbf{53}, ~1--7 (2014)

\bibitem{ding2018empirical}
Ding, R., Li, X., Nie, L., Li, J., Si, X., Chu, D., Liu, G., Zhan, D.: Empirical study and improvement on deep transfer learning for human activity recognition. Sensors  \textbf{19}(1), ~57 (2018)

\bibitem{esling2012time}
Esling, P., Agon, C.: Time-series data mining. ACM Computing Surveys (CSUR)  \textbf{45}(1),  1--34 (2012)

\bibitem{fernando2013unsupervised}
Fernando, B., Habrard, A., Sebban, M., Tuytelaars, T.: Unsupervised visual domain adaptation using subspace alignment. In: Proceedings of the IEEE international conference on computer vision. pp. 2960--2967 (2013)

\bibitem{flamary2016optimal}
Flamary, R., Courty, N., Tuia, D., Rakotomamonjy, A.: Optimal transport for domain adaptation. IEEE Trans. Pattern Anal. Mach. Intell  \textbf{1},  1--40 (2016)

\bibitem{guo2003knn}
Guo, G., Wang, H., Bell, D., Bi, Y., Greer, K.: Knn model-based approach in classification. In: On The Move to Meaningful Internet Systems 2003: CoopIS, DOA, and ODBASE: OTM Confederated International Conferences, CoopIS, DOA, and ODBASE 2003, Catania, Sicily, Italy, November 3-7, 2003. Proceedings. pp. 986--996. Springer (2003)

\bibitem{kerdoncuff2021metric}
Kerdoncuff, T., Emonet, R., Sebban, M.: Metric learning in optimal transport for domain adaptation. In: Proceedings of the Twenty-Ninth International Conference on International Joint Conferences on Artificial Intelligence. pp. 2162--2168 (2021)

\bibitem{lara2012survey}
Lara, O.D., Labrador, M.A.: A survey on human activity recognition using wearable sensors. IEEE communications surveys \& tutorials  \textbf{15}(3),  1192--1209 (2012)

\bibitem{lentzas2020non}
Lentzas, A., Vrakas, D.: Non-intrusive human activity recognition and abnormal behavior detection on elderly people: A review. Artificial Intelligence Review  \textbf{53}(3),  1975--2021 (2020)

\bibitem{li2020simultaneous}
Li, S., Xie, B., Wu, J., Zhao, Y., Liu, C.H., Ding, Z.: Simultaneous semantic alignment network for heterogeneous domain adaptation. In: Proceedings of the 28th ACM international conference on multimedia. pp. 3866--3874 (2020)

\bibitem{lu2018learning}
Lu, J., Liu, A., Dong, F., Gu, F., Gama, J., Zhang, G.: Learning under concept drift: A review. IEEE transactions on knowledge and data engineering  \textbf{31}(12),  2346--2363 (2018)

\bibitem{lu2021cross}
Lu, W., Chen, Y., Wang, J., Qin, X.: Cross-domain activity recognition via substructural optimal transport. Neurocomputing  \textbf{454},  65--75 (2021)

\bibitem{pan2009survey}
Pan, S.J., Yang, Q.: A survey on transfer learning. IEEE Transactions on knowledge and data engineering  \textbf{22}(10),  1345--1359 (2009)

\bibitem{patel2015visual}
Patel, V.M., Gopalan, R., Li, R., Chellappa, R.: Visual domain adaptation: A survey of recent advances. IEEE signal processing magazine  \textbf{32}(3),  53--69 (2015)

\bibitem{rabiner1986introduction}
Rabiner, L., Juang, B.: An introduction to hidden markov models. ieee assp magazine  \textbf{3}(1),  4--16 (1986)

\bibitem{reiss2012introducing}
Reiss, A., Stricker, D.: Introducing a new benchmarked dataset for activity monitoring. In: 2012 16th international symposium on wearable computers. pp. 108--109. IEEE (2012)

\bibitem{rokni2018autonomous}
Rokni, S.A., Ghasemzadeh, H.: Autonomous training of activity recognition algorithms in mobile sensors: A transfer learning approach in context-invariant views. IEEE Transactions on Mobile Computing  \textbf{17}(8),  1764--1777 (2018)

\bibitem{saeedi2018personalized}
Saeedi, R., Sasani, K., Norgaard, S., Gebremedhin, A.H.: Personalized human activity recognition using wearables: A manifold learning-based knowledge transfer. In: 2018 40th Annual International Conference of the IEEE Engineering in Medicine and Biology Society (EMBC). pp. 1193--1196. IEEE (2018)

\bibitem{sekiguchi2020ensemble}
Sekiguchi, R., Abe, K., Yokoyama, T., Kumano, M., Kawakatsu, M.: Ensemble learning for human activity recognition. In: Adjunct proceedings of the 2020 ACM international joint conference on pervasive and ubiquitous computing and proceedings of the 2020 ACM international symposium on wearable computers. pp. 335--339 (2020)

\bibitem{soleimani2021cross}
Soleimani, E., Nazerfard, E.: Cross-subject transfer learning in human activity recognition systems using generative adversarial networks. Neurocomputing  \textbf{426},  26--34 (2021)

\bibitem{sukhija2019supervised}
Sukhija, S., Krishnan, N.C.: Supervised heterogeneous feature transfer via random forests. Artificial Intelligence  \textbf{268},  30--53 (2019)

\bibitem{sun2016return}
Sun, B., Feng, J., Saenko, K.: Return of frustratingly easy domain adaptation. In: Proceedings of the AAAI conference on artificial intelligence. vol.~30 (2016)

\bibitem{tan2018survey}
Tan, C., Sun, F., Kong, T., Zhang, W., Yang, C., Liu, C.: A survey on deep transfer learning. In: Artificial Neural Networks and Machine Learning--ICANN 2018: 27th International Conference on Artificial Neural Networks, Rhodes, Greece, October 4-7, 2018, Proceedings, Part III 27. pp. 270--279. Springer (2018)

\bibitem{wang2018impact}
Wang, G., Li, Q., Wang, L., Wang, W., Wu, M., Liu, T.: Impact of sliding window length in indoor human motion modes and pose pattern recognition based on smartphone sensors. Sensors  \textbf{18}(6), ~1965 (2018)

\bibitem{wang2018device}
Wang, J., Gao, Q., Pan, M., Fang, Y.: Device-free wireless sensing: Challenges, opportunities, and applications. IEEE network  \textbf{32}(2),  132--137 (2018)

\bibitem{wang2018stratified}
Wang, J., Chen, Y., Hu, L., Peng, X., Philip, S.Y.: Stratified transfer learning for cross-domain activity recognition. In: 2018 IEEE international conference on pervasive computing and communications (PerCom). pp. 1--10. IEEE (2018)

\bibitem{weinberger2009distance}
Weinberger, K.Q., Saul, L.K.: Distance metric learning for large margin nearest neighbor classification. Journal of machine learning research  \textbf{10}(2) (2009)

\bibitem{wilson2020survey}
Wilson, G., Cook, D.J.: A survey of unsupervised deep domain adaptation. ACM Transactions on Intelligent Systems and Technology (TIST)  \textbf{11}(5),  1--46 (2020)

\bibitem{xing2018enabling}
Xing, T., Sandha, S.S., Balaji, B., Chakraborty, S., Srivastava, M.: Enabling edge devices that learn from each other: Cross modal training for activity recognition. In: Proceedings of the 1st International Workshop on Edge Systems, Analytics and Networking. pp. 37--42 (2018)

\bibitem{zhao2011cross}
Zhao, Z., Chen, Y., Liu, J., Shen, Z., Liu, M.: Cross-people mobile-phone based activity recognition. In: Twenty-second international joint conference on artificial intelligence. Citeseer (2011)

\end{thebibliography}

\end{document}